\PassOptionsToPackage{usenames, dvipsnames}{xcolor}
\documentclass[sigconf,nonacm]{acmart}

\usepackage{algorithmic}
\usepackage{graphicx}
\usepackage{textcomp}
\usepackage{balance}
\usepackage[normalem]{ulem}
\usepackage{lipsum}
\usepackage{booktabs}
\usepackage{mathtools}
\usepackage[english]{babel}
\usepackage{multirow}
\usepackage{subfig}
\usepackage{bbm}
\usepackage{url}
\usepackage{hyperref}
\usepackage{amsmath}
\usepackage{amsthm}
\usepackage{stmaryrd}
\usepackage{listings}
\usepackage{dsfont}
\usepackage{longtable}
\usepackage{lscape}
\usepackage{blindtext}
\usepackage{enumitem}
\setlist[enumerate]{leftmargin=*}

\usepackage[export]{adjustbox}
\usepackage[english,ruled,vlined,linesnumbered]{algorithm2e}
\usepackage[justification=centering]{caption}

\theoremstyle{definition}

\newtheorem{problem}{Problem}

\newcommand{\mycomment}[1]{}

{\renewcommand{\arraystretch}{1.5}}

\AtBeginDocument{
  \catcode`_=12
  \begingroup\lccode`~=`_
  \lowercase{\endgroup\let~}\sb
  \mathcode`_="8000
}

\DeclareMathOperator*{\argmax}{arg\,max}
\DeclareMathOperator*{\argmin}{arg\,min}

\let\origthelstnumber\thelstnumber
\makeatletter
\newcommand*\Suppressnumber{%
  \lst@AddToHook{OnNewLine}{%
    \let\thelstnumber\relax%
     \advance\c@lstnumber-\@ne\relax%
    }%
}

\definecolor{codegreen}{rgb}{0,0.6,0}
\definecolor{codegray}{rgb}{0.5,0.5,0.5}
\definecolor{codepurple}{rgb}{0.58,0,0.82}
\definecolor{backcolour}{rgb}{0.95,0.95,0.92}

\newcommand*\Reactivatenumber[1]{%
  \setcounter{lstnumber}{\numexpr#1-1\relax}
  \lst@AddToHook{OnNewLine}{%
   \let\thelstnumber\origthelstnumber%
   \refstepcounter{lstnumber}
  }%
}

\usepackage{fancyvrb}

\begin{document}

\title{DeepLSH: Deep Locality-Sensitive Hash Learning for Fast and Efficient Near-Duplicate Crash Report Detection} 

\author{Youcef Remil}
\affiliation{%
  \institution{INSA Lyon} 
  \institution{Infologic R\&D}
  \city{26500 Bourg-Lès-Valence}
  \country{France}
}
\email{yre@infologic.fr}

\author{Anes Bendimerad}
\affiliation{%
  \institution{Infologic R\&D}
  \city{26500 Bourg-Lès-Valence}
  \country{France}
}
\email{abe@infologic.fr}

\author{Romain Mathonat}
\affiliation{%
  \institution{Infologic R\&D}
  \city{26500 Bourg-Lès-Valence}
  \country{France}
}
\email{rma@infologic.fr}

\author{Chedy Raissi}
\affiliation{%
  \institution{Riot Games}
  \city{018937 Marina One West Tower}
  \country{Singapore}
}
\email{chedy.raissi@inria.fr}

\author{Mehdi Kaytoue}
\affiliation{%
  \institution{INSA Lyon} 
  \institution{Infologic R\&D}
  \city{26500 Bourg-Lès-Valence}
  \country{France}
}
\email{mka@infologic.fr}

\begin{abstract}
Automatic crash bucketing is a crucial phase in the software development process for efficiently triaging bug reports. It generally consists in grouping similar reports through clustering techniques. However, with real-time streaming bug collection, systems are needed to quickly answer the question: \textit{What are the most similar bugs to a new one?}, that is, efficiently find near-duplicates. It is thus natural to consider nearest neighbors search to tackle this problem and especially the well-known locality-sensitive hashing (LSH) to deal with large datasets due to its sublinear performance and theoretical guarantees on the similarity search accuracy. Surprisingly, LSH has not been considered in the crash bucketing literature. It is indeed not trivial to derive hash functions that satisfy the so-called \textit{locality-sensitive property} for the most advanced crash bucketing metrics. Consequently, we study in this paper how to leverage LSH for this task. To be able to consider the most relevant metrics used in the literature, we introduce {\sc DeepLSH}, a Siamese DNN architecture with an original loss function, that perfectly approximates the locality-sensitivity property even for Jaccard and Cosine metrics for which exact LSH solutions exist. We support this claim with a series of experiments on an original dataset, which we make available.

\end{abstract}



\maketitle

\section{Introduction}\label{sec:introduction}

\mycomment{
\begin{figure} \small
\begin{Verbatim}[frame=single]
id: 16377610978254995717215-XXXXXXX-XX
sessionId: 2D7E2416131887D473F6CFD7B35769C
version: 13.7
server: XXXXXXX-XX
timestamp: 2022-07-26 11:13:40.657
typeError: ERROR
functionality: com.company.modules.factory.Factory
message:  No CAB matches reading 'Invalid' 
detail: class com.company.exceptions.MyException: 
    at com.company.LancAdapter.do(LancAdapter.java:449)
    at com.company.CABWrapper.read(CABWrapper.java:191)
    ...
    at com.company.Main(Main.java:94)
User message : I got this error while I was trying
    to order a purchase [...]
\end{Verbatim}
\caption{  A crash report with a stack trace and its context. User message is only available for user-generated crash reports.}\label{fig:crashexample}
\end{figure}}

Collection and triage of runtime errors following a software release are integral parts of a standard quality process (e.g., Windows Error Reporting~\cite{DBLP:conf/sosp/GlerumKGAONGLH09}, Mozilla Crash Reporter~\cite{website:mozillaCR}). At our company, which specializes in editing and installing b2b Enterprise Resource Planning (ERP) systems, we receive a daily influx of over ten thousand of automatic and user-generated reports. Each report comprises a Java stack trace and relevant contextual information. It is important to note that not all crash reports hold the same level of priority: Rare occurrences indicate unexpected shutdowns, more frequent ones involve GUI issues that obstruct end-users from completing their tasks, while the majority fall into the "silent" category. The latter typically represents bugs that either have workarounds found by users (thus limiting their impact) or background process failures that may take days to notice but have significant consequences. Hence, it is imperative to promptly address such bugs.

While extreme problems are rare and easy to identify and prioritize, it remains challenging to sort, rank and assign other reports to developers (or simply ignore them). Indeed, many different reports may actually imply a single root cause and a bug can produce slightly different stack traces known as \textit{near-duplicates}~\cite{DBLP:conf/icse/DangWZZN12}. Therefore, it is highly valuable to group similar crashes into buckets to accelerate the crash investigation process~\cite{DBLP:conf/icsm/DhaliwalKZ11}. Reporting systems use a range of heuristics and manually developed rules to organize crash reports into categories, ideally each referring to the same bug~\cite{DBLP:conf/sosp/GlerumKGAONGLH09}. However, in many cases, it may assign crash reports caused by the same bug to multiple buckets~\cite{DBLP:conf/sosp/GlerumKGAONGLH09}. Thus, various alternatives address the \textit{stack trace-based report deduplication problem} by designing custom and accurate similarity measures between stack traces, relying mostly on string and graph matching (e.g., edit distance, prefix match and LCSS)~\cite{DBLP:conf/icsm/DhaliwalKZ11,DBLP:conf/dsn/KimZN11, DBLP:conf/icac/BrodieMLMMWCS05} and information retrieval (e.g., TF-IDF and N-grams)~\cite{DBLP:conf/csmr/LerchM13,DBLP:conf/qrs/SaborHL17}.  Other metrics take into account specific characteristics of stack traces, such as the distance to the top frame or alignment between matched frames)~\cite{DBLP:conf/sigsoft/VasilievKCKLP20, DBLP:conf/icse/DangWZZN12, DBLP:conf/seke/MorooAH17}. In the aforementioned studies, similarity measures are generally embedded in clustering algorithms, but this comes with several drawbacks: First, it needs numerous similarity calculations when assigning a new stack trace to a cluster. Second, clusters are not stable over time and should be recalculated frequently without losing links to actual bug tickets that have been previously created. It can also be difficult to set the various parameters that need to be tuned. 


In our company, we aim to process reports in quasi-real time and enhance our maintenance processes. An essential aspect of this objective is efficiently identifying the nearest neighbors for each crash report. We need to quickly determine if a corresponding ticket has already been created, allowing us to update its statistics, or if a new ticket needs to be generated. This becomes even more crucial when end-users directly contact us via email, phone, or assistance tickets. We must perform the same checks to ascertain whether similar issues have been encountered before and if any temporary or permanent solutions have been provided. However, the computational demands of searching for nearest neighbors are generally prohibitive. For instance, we faced a significant challenge when conducting linear scans, which took approximately 10 hours to compare 1,000 stack traces against a pool of 100,000 stack traces using the similarity metric proposed by Brodie et al.\cite{DBLP:conf/icac/BrodieMLMMWCS05}. This well-known observation within the field of data mining has led us to explore the concept of \textit{approximate nearest neighbors search} (ANN)\cite{DBLP:journals/corr/abs-2003-03369}.

Hashing is a popular technique for ANN ~\cite{DBLP:journals/corr/abs-2003-03369}, and particularly LSH, allowing to search for approximate nearest neighbors in constant time. LSH satisfies the \textit{locality-sensitive property}, that is, similar items are expected to have a higher probability to be mapped into the same hash code (or hash bucket) than dissimilar items \cite{DBLP:journals/corr/WangSSJ14}. Most importantly, LSH provides guarantees on the search accuracy that is, \textit{the probability for two stack traces having randomly the same hash function is equal to their similarity} and the collision probability of being hashed into at least one bucket can be simply and fully controlled with two key parameters representing the user's desired search precision and recall. The more accurate LSH is, the larger its hash tables are. Fortunately, LSH is known for its computational and storage efficiency, as well as its sublinear search performance~\cite{DBLP:conf/stoc/IndykM98}.

LSH remains surprisingly unexplored in the crash bucketing literature, despite its potential benefits. It is indeed not trivial to derive hash functions that guarantee the \textit{locality-sensitive property} for the most advanced metrics of crash bucketing. Generating hash functions that meet this property is a non-trivial task. Although several LSH function families have been proposed, each for estimating one and only one conventional similarity/distance measure, e.g., Min-Hash function for Jaccard coefficient~\cite{DBLP:journals/cn/BroderGMZ97} and Sign-Random-Projection (Sim-Hash) function for angular distance~\cite{DBLP:conf/stoc/Charikar02}, etc., a generalized procedure for applying LSH to various similarity measures remains ambiguous and theoretically complex. More specifically, there is currently no systematic procedure for deriving a family of LSH functions for any given similarity measure.

Exploring the application of LSH to custom similarities for crash deduplication is a novel area that we delve into. We propose to learn these hash functions in a supervised manner to mimic any given similarity measure while incorporating the locality-sensitive hashing component into the model learning process. We draw inspiration from the field of \textit{learn to hash} techniques~\cite{DBLP:conf/cvpr/LiongLWMZ15,DBLP:conf/cvpr/Liu0SC16,DBLP:conf/nips/LiSHT17,DBLP:conf/eccv/DoDC16,DBLP:conf/cvpr/LaiPLY15,DBLP:conf/accv/WangSK16}, which effectively reduce the dimensionality of input data representations while preserving similarity. In fact, we aim to leverage the strengths of both LSH and Learn to Hash approaches. Our proposed model generates a hash code, with LSH guarantees, that is shared by the nearest neighbors of the input stack trace. To the best of our knowledge, this is the first \textit{similarity-agnostic} method that utilizes hashing for the crash deduplication problem. It is important to note that our objective is not to introduce a new similarity measure, but rather to enable existing measures to scale effectively.

\noindent \textbf{Contribution.} Our contribution is three-fold. (i) Aiming to overcome the problem of deriving LSH functions for stack-trace similarity measures, we propose a generic approach dubbed {\sc DeepLSH} that learns and provides a family of binary hash functions that perfectly approximate the \textit{locality-sensitive property} to retrieve efficiently and rapidly near-duplicate stack traces. (ii) Technically, we design a deep Siamese neural network architecture to perform end-to-end hashing with an original objective loss function based on the locality-sensitive property preserving with appropriate regularizations to cope with the binarization problem of optimizing non-smooth loss functions. (iii) We demonstrate through our experimental study the effectiveness and scalability of {\sc DeepLSH} to yield near-duplicate crash reports under a dozen of similarity metrics. We successfully compare to standard LSH techniques (MinHash and SimHash), and the most relevant deep hashing baseline~\cite{DBLP:journals/tip/HuFXSCM18} on a large real-world dataset that we make available.

\noindent \textbf{Main findings.} (i) {\sc DeepLSH} demonstrates exceptional convergence to the locality sensitive property for all studied stack-trace similarity measures: it effectively maintains the LSH guarantees, exhibiting high precision/recall. (ii) {\sc DeepLSH}  consistently achieves satisfactory search accuracy, with a recall rate and a ranking quality up to 0.9 for unseen stack traces. (iii) {\sc DeepLSH} almost matches MinHash (for Jaccard) and outperforms SimHash (for Cosine) in search performance while still generalizing to other complex similarity metrics. (iv) {\sc DeepLSH} is highly scalable and can retrieve near-duplicate crash reports in a constant time (an average of 24 seconds on 100 millions queries) compared to an exact K-NN approach which takes hours to perform a linear scan. (v) Our end-to-end hashing approach combining LSH and learn-to-hash has shown to be significantly more accurate than using only learn-to-hash or LSH performed separately on learn to hash results, as demonstrated by comparison to the most identified relevant baseline~\cite{DBLP:journals/tip/HuFXSCM18}.  

\section{Related work}\label{sec:relatedwork}

Our work encompasses two research areas which will be discussed in this section: (1) the approximate nearest neighbor search through hashing techniques, and (2) custom similarity measures for the stack trace deduplication problem.

\noindent \textbf{Hashing for ANN search.} Locality-sensitive hashing has been widely studied by the theoretical computer science community. Its main aspect focuses on the generation of a family of random hash functions that meet the locality-sensitive property for conventional similarity measures~\cite{sadowski2007simhash, DBLP:conf/compgeom/DatarIIM04,DBLP:conf/stoc/Charikar02,DBLP:conf/stoc/IndykM98,DBLP:journals/cn/BroderGMZ97}. 
Particularly, Min-Hash (or min-wise independent permutations)~\cite{DBLP:journals/cn/BroderGMZ97} is an LSH function designed specifically for Jaccard similarity. Sim-Hash~\cite{DBLP:conf/stoc/Charikar02,sadowski2007simhash} is another popular technique whose aim is to estimate angular similarities such as Cosine. Sim-Hash has been adopted by Google~\cite{sadowski2007simhash} and it is often used in text processing applications to compare between documents.
Both techniques cannot, however, be applied to estimate other similarity metrics besides Jaccard or Cosine. LSH has garnered limited interest within the software engineering community, mainly being solicited for tasks like code search~\cite{silavong2022senatus} and clone detection~\cite{jiang2007deckard} through the utilization of well-known LSH functions such as MinHash and Hamming LSH. However, it remains an unexplored area in the domain of crash-deduplication, primarily due to the fact that the currently used metrics do not facilitate the application of conventional and well-established LSH functions. 

Designing LSH functions for any given similarity metric remains ambiguous and theoretically challenging, as there is no established method for deriving a set of LSH hash functions for a specific similarity measure. On the other hand, the concept of learn to hash has become the focus of many learning-based hashing methods especially for the computer vision community~\cite{DBLP:conf/cvpr/LiongLWMZ15,DBLP:conf/cvpr/Liu0SC16,DBLP:journals/tip/HuFXSCM18,DBLP:conf/cvpr/0003CBS18, DBLP:conf/cvpr/LaiPLY15,DBLP:conf/accv/WangSK16, DBLP:conf/cvpr/LinYHC15,DBLP:conf/cvpr/ZhangCS16}. These methods are primarily designed for searching image similarity and have proven to be highly effective in reducing the dimensionality of input data representations while preserving their similarities. However, they do not meet our main objective, which consists in an end-to-end procedure to retrieve near-duplicate data objects with guarantees. They do not reveal a systematic way to construct hash tables from the resulting hash codes, and neither do they control the trade-off between recall and precision using key parameters as does LSH. Alternatively, we take benefit from both worlds, i.e., LSH and Learn to Hash, by proposing an end-to-end procedure that incorporates the LSH component in the learning process. We have identified a similar baseline approach in~\cite{DBLP:journals/tip/HuFXSCM18} that proposes a different methodology compared to ours, consisting of a deep hash coding neural network combined with Hamming LSH fitted on the resulting hash vectors to retrieve near-duplicate images in a large database. The authors first proposed a constrained loss function without incorporating the locality-sensitive property, and then performs discretization on continuous hash vectors to carry out the Hamming LSH separately from the model. We show through our experiments in Sec.~\ref{sec:xp} by adapting this two-step approach on stack traces, that it leads to a considerable search performance degradation compared to our approach {\sc DeepLSH}.

\noindent \textbf{Stack trace similarities.} 
We report research studies that tackle the crash report deduplication problem using stack trace similarity functions. Lerch and Mezini~\cite{DBLP:conf/csmr/LerchM13} employed the TF-IDF-based scoring function from Lucene library~\cite{website:lucene}. Sabor et al.~\cite{DBLP:conf/qrs/SaborHL17} proposed DURFEX system which uses the package name of the subroutines and then segment the resulting stack traces into N-grams to compare them using the Cosine similarity. Some alternative techniques propose to compute the similarity using derivatives of the Needleman-Wunsch algorithm~\cite{needleman1970general}. In~\cite{DBLP:conf/icac/BrodieMLMMWCS05}, Brodie et al. suggested to adjust the similarity based on the frequency and the position of the matched subroutines. Dang et al.~\cite{DBLP:conf/icse/DangWZZN12} proposed a new similarity measure called PDM in their framework Rebucket to compute the similarity based on the offset distance between the matched frames
and the distance to the top frame. More recently, TraceSim~\cite{DBLP:conf/sigsoft/VasilievKCKLP20} has been proposed to take into consideration both the frame position and its global inverse frequency. Moroo et al~\cite{DBLP:conf/seke/MorooAH17} present an approach that combines TF-IDF coefficient with PDM. Finally we outline some earlier approaches that used edit distance as it is equivalent to optimal global alignment~\cite{DBLP:conf/osdi/BartzSPKGCL08,DBLP:conf/icde/ModaniGLSM07}. Note that our approach {\sc DeepLSH} does not propose a new similarity measure and does not question the effectiveness or compete against these existing measures, but it is complementary to them. We demonstrate that {\sc DeepLSH} model is able to estimate all these measures with the purpose of providing a scalable way to yield approximate near-duplicate stack traces w.r.t these custom similarity functions.
\section{Background and Problem definition}\label{sec:background} 

\subsection{Crash reports and stack-trace dataset}
Software often contains bugs that can lead to crashes and errors. In the following, we use both terms interchangeably to refer to instances of application crashes, where the system becomes unresponsive, as well as errors arising from background tasks or error pop-ups presented to end-users. Each of these issues is accompanied by a Java stack trace and a run-time context (software/OS/database version, timestamp, etc.) \cite{DBLP:conf/msr/SchroterBP10}. A stack trace is a detailed report of the executed methods and their associated packages during a crash. Stack traces can be retrieved through system calls in many programming languages. In Java, the stack trace lists methods in descending order, with the top of the stack trace representing the most inner call. This is illustrated with a crash report from a software product in Fig.~\ref{fig:crashexample}. We define the stack-trace dataset as set of $N$ stack traces $\mathcal{D} = \{s_1, s_2, ..., s_N \}$.

\begin{figure}[!h]
\centering
    \begin{lstlisting} 
        id: 16377610978254995717215-XXXXXXX-XX
        sessionId: 2D7E2416131887D473F6CFD7B35769C
        version: 13.7
        @timestamp: 2022-12-26 11:13:40.657
        typeError: ERROR
        functionality: com.company.modules.factory.Factory
        message:  No CAB matches reading 'Invalid' 
        detail: class com.company.exceptions.MyException: 
            at com.company.LancAdapter.do(LancAdapter.java:449)
            at com.company.CABWrapper.read(CABWrapper.java:191)
            ...
            at com.company.Main(Main.java:94)
        user message : I got this error while I was trying to ...
    \end{lstlisting}
    \caption{A crash report with a stack trace and its context.}\label{fig:crashexample}
\end{figure}

\subsection{Approximate Nearest Neighbors Search}
A similarity measure between two stack traces is a function denoted as $sim : \mathcal{D} \times \mathcal{D} \longrightarrow [0,1]$. 
It can be any conventional similarity metric (Jaccard coefficient) or a specialized stack-trace similarity measure (e.g., PDM~\cite{DBLP:conf/icse/DangWZZN12} and TraceSim~\cite{DBLP:conf/sigsoft/VasilievKCKLP20}). The distance function is naturally given by $dist : \mathcal{D} \times \mathcal{D} \longrightarrow [0,1]$ where $dist = 1-sim$. 
Given a dataset $\mathcal{D}$ of $N$ stack traces, the problem of nearest neighbor search under a user-defined similarity measure $sim$ consists in finding, for a specific stack trace $s \in \mathcal{D}$, another stack trace denoted as $nn(s) \in \mathcal{D} \setminus \{ s\}$ such that: 
$nn(s) = \argmax_{s' \in \mathcal{D} \setminus \{ s\}} sim(s,s')$.

An alternative of nearest neighbor search is the fixed-radius nearest neighbor ($R-$near neighbor) problem which seeks to find a set of stack traces $\mathcal{S}_R$ that are within the distance $R$ of $s$ $(0<R<1)$, such that: $\mathcal{S}_R = \{s' \in \mathcal{D} \setminus \{ s\} \mid  dist(s,s') \leq R\}$. 

There exists simple tree-based algorithms for approximate nearest neighbor search problems, notably KD trees~\cite{DBLP:journals/cacm/Bentley75} and SR-tree~\cite{DBLP:conf/sigmod/KatayamaS97}. However, for large scale high-dimensional cases, these techniques suffer from the well-known  \textit{curse of dimensionality}~\cite{DBLP:conf/icdt/BeyerGRS99} where the performance is often surpassed by a linear scan. Consequently, significant research efforts have been dedicated to exploring highly efficient and scalable methods for approximating nearest neighbor search problems in large-scale datasets, including hashing and LSH. 

Locality-Sensitive Hashing (LSH) has been particularly proposed to tackle the problem of \textit{randomized or probabilistic approximate nearest neighbors search}~\cite{DBLP:journals/corr/WangSSJ14}, that is, targeting the ANN problem with guarantees aiming to find approximate nearest neighbors with probability rather than a deterministic way (which is not tractable). This choice is driven by the purpose of ensuring guarantees on the search accuracy with respect to the exact nearest neighbors search, while giving the user the ability to balance between precision and recall to a desired level. Formally, we define our problem as follows:

\begin{problem}[Randomized approximate nearest neighbors search (RANN)] \label{problem}
Given a new reported stack trace $s$, a dataset $\mathcal{D}$  of historical stack traces, the goal is to report some of the $R-$nearest neighbors $\mathcal{R}$ of $s$ such that: $\mathcal{R} = \{s' \in \mathcal{D} \setminus \{ s\} \mid  Pr[s' \in \mathcal{S}_R] \geq 1-\delta \}$ with $(0<\delta<1)$. The lower the parameter $\delta$, the lower the chance of finding elements in the radius (i.e., more restrictive).
\end{problem}

\subsection{Hashing approach for the RANN problem}
Hashing-based approaches attempt to map data features from the input space into a lower-dimensional space using hash functions so that the approximate nearest neighbors search on the resulting hash vectors can be performed efficiently.
The compact hash codes generally belong to the Hamming space i.e., binary codes. We define the hash function for a stack trace $s$ as $y = h(s)$ where $y$ is the hash code and $h : \mathcal{D} \longrightarrow \{0,1\}^b$ where $b \geq 1$ is the number of bits in the hash code. In approximate nearest neighbors search settings, we usually opt for multiple hash functions to compute the final meta-hash code: $Y = H(s)$, where $H(s) = [h_1(s), h_2(s), ..., h_K(s)]^T$ and $K$ is the number of hash functions. Hashing-based nearest neighbors search includes hash table lookup strategy~\cite{DBLP:journals/corr/WangSSJ14} which seeks to design an efficient search scheme rooted in hash tables. The hash table is a data structure made of buckets, each of which is indexed by a meta-hash code such that the probability of collision of near-duplicate stack traces under a given similarity measure is maximized.
Given a stack trace $s$, the stack traces $\{s' \in \mathcal{D} \setminus \{ s\} \mid H(s) = H(s')\}$ are retrieved as near-duplicates of $s$. In order to improve the recall, we generally construct $L$ hash tables containing hash buckets, each corresponding to a hash code $\{H_1, H_2, ..., H_L\}$. The near-duplicate stack traces are then defined as $\{s' \in \mathcal{D} \setminus \{ s\} \mid  \exists j \in [\![1,L]\!], H_j(s) = H_j(s') \}$.

\subsection{LSH for RANN problem}\label{subsec:lsh}

To address the problem~\ref{problem} of randomized nearest neighbors search, Locality-Sensitive Hashing (LSH)~\cite{DBLP:conf/stoc/IndykM98} maps high dimensional data to lower dimensional representations by using a family $\mathcal{H}$ of random hash functions that satisfy the locality-sensitive property.
Thus, similar data items  in the high-dimensional input space are expected to have more chance to be mapped to the same hash buckets than dissimilar items.
These similar data items are said to collide. 
Starting with a formal definition of an LSH family $\mathcal{H}$ to address our problem, we consider a metric space such that, $\mathcal{M} = (\mathcal{D}, dist)$, 
a threshold $0<R<1$, an approximation factor $c > 1$, and two probabilities $p_1$ and $p_2$. The hash family $\mathcal{H}$ is a set of $M$ hash functions $\{h_1, h_2, ..., h_M\}$ where each $h \in \mathcal{H}$ is defined as $h : \mathcal{D} \longrightarrow \{0,1\}^b$. An LSH family must satisfy the following conditions for any two stack traces $s, s' \in \mathcal{D}$ and any random hash function $h \in \mathcal{H}$:
\begin{itemize}
    \item if $dist(s,s') \leq R$,  then $Pr[h(s)=h(s')] \geq p_1$, 
    \item if $dist(s,s') \geq cR$, then $Pr[h(s)=h(s')] \leq p_2$.
\end{itemize}

A family $\mathcal{H}$ is said to be $(R, cR, p_1, p_2)-$sensitive if $p_1 > p_2$. Alternatively~\cite{DBLP:conf/stoc/Charikar02}, a sufficient condition for $\mathcal{H}$ to be an LSH family is that the \textit{collision probability} should be monotonically increasing with the similarity i.e.,
{
\begin{equation}
    Pr[h(s)=h(s')] = g(sim(s,s')),
\label{eq:collisionProb}
\end{equation}}
where $g$ is a monotonically increasing function. Indeed, most of popular known LSH families such as Minhash~\cite{DBLP:journals/cn/BroderGMZ97} for Jaccard similarity, satisfy this strong property.

\begin{figure}
\centering
 \includegraphics[width=0.5\textwidth]{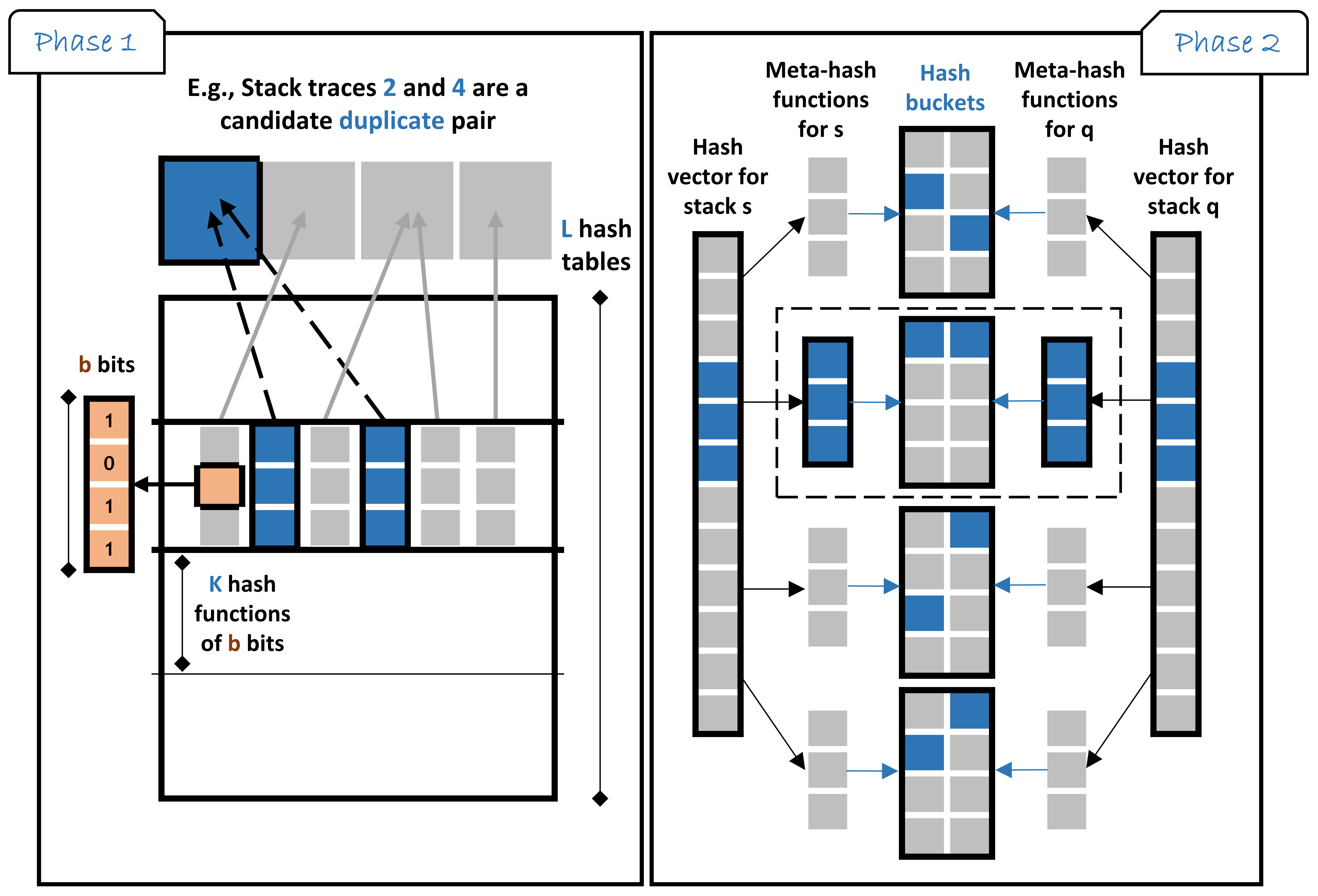}
\caption{\label{fig:lshOverview}$(L,K)-$ parameterized LSH algorithm for retrieving near-duplicate stack traces with guarantees.}
\end{figure}

The LSH scheme indexes all stack traces in hash tables and searches for near-duplicates via a hash table lookup strategy. The LSH algorithm uses two key hyperparameters $L$ and $K$ to be tuned. Given the LSH family $\mathcal{H}$, the LSH algorithm amplifies the gap between the high probability $p_1$ and the low probability $p_2$ by concatenating $K$ hash functions chosen independently and uniformly at random from $\mathcal{H}$, to form a meta-hash function $H(s) = [h_1(s), h_2(s), ..., h_K(s)]^T$. The meta-hash function is associated with a bucket ID in a hash table. Intuitively, it reduces the chances of collision between similar stack traces, since this requires them to have the same value for each of the $K$ hash functions (i.e., high precision over the recall). To improve the recall, $L$ meta-hash functions $H_1, H_2, ..., H_L$ are sampled independently, each of which corresponds to a hash table. These meta-hash functions are used to map each stack trace into $L$ hash codes, and $L$ hash tables are constructed to index the corresponding buckets, each using $K$ random hash functions. The LSH algorithm is conducted in two phases as illustrated in Fig.~\ref{fig:lshOverview} (considering, $L=4$ hash tables, for each we have $K=3$ hash functions of $b=4$ bits.) 

\smallbreak
\noindent \textbf{Pre-processing phase:} The $L$ hash tables are built from $N$ stack traces. A hash table is indexed by $K$ hash functions constituting its meta-hash function. We store pointers to stack traces in hash tables, since storing them in the original format is memory intensive.

\smallbreak
\noindent \textbf{Querying phase:} Given a stack trace $s$, the algorithm iterates over the $L$ meta-hash functions in order to retrieve all stack traces that are hashed into the same bucket as $s$, then reports the union from all these buckets $\bigcup_{j=1}^L \{s' \in \mathcal{D} \mid  H_j(s) = H_j(s')\}$. A $(L,K)-$parameterized LSH algorithm succeeds in finding candidate near-duplicates for a stack trace $s$ with a sampling probability at least $1-(1-p^K)^L$, where $p$ is the collision probability of LSH function. This means that $\delta = (1-p^K)^L$ as defined in the problem~\ref{problem} of randomized ANN. If property~\eqref{eq:collisionProb} holds, in particular for the identity function, i.e., $g(x) = I_x$, we can rely on the so-called \textit{probability-similarity} relation between two different stack traces $s, s' \in \mathcal{D}$ such that:
    \begin{equation}
        P_{K,L} (s,s')=1-(1-sim(s,s')^K)^L 
    \label{eq:relationProbSim}
    \end{equation}

\section{DeepLSH Design Methodology}\label{sec:methodology}

\begin{figure*}[!t]
\centering
\includegraphics[width=0.9\textwidth]{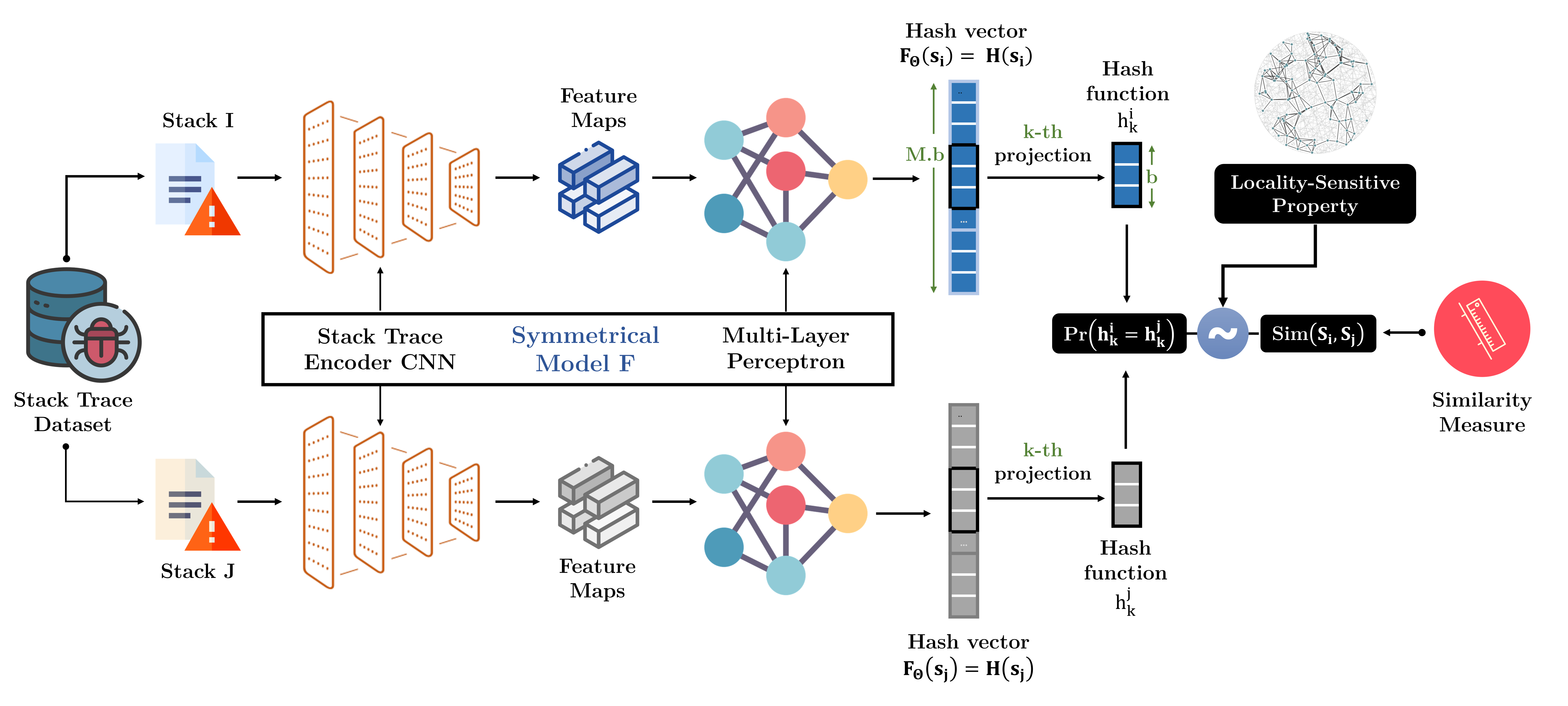}
\caption{\label{fig:siamese1} {\sc DeepLSH}: Deep Siamese hash learning neural network overview}
\end{figure*}

In order to address Problem~\ref{problem} to efficiently retrieve near-duplicates, it is crucial to define suitable Locality-Sensitive Hashing (LSH) families for stack trace-based similarity measures. While several LSH families have been proposed for various similarity measures~\cite{DBLP:conf/compgeom/DatarIIM04,DBLP:conf/stoc/Charikar02,DBLP:journals/cn/BroderGMZ97}, there is currently no generic mechanism available to generate a family of hash functions that satisfies the locality-sensitive property for any user-defined similarity measure, especially non-linear measures that often require human expertise. To overcome this challenge, we introduce {\sc DeepLSH}, a generic approach that solely requires the stack trace dataset and any user-defined similarity measure (measure-agnostic) as input. {\sc DeepLSH} provides a family of hash functions that converges to the locality-sensitive property.

\subsection{Learning a family of LSH functions}\label{subsec:learnLSH}

We exploit a deep supervised Siamese neural network with an original objective loss function to learn hash functions that converge to the locality-sensitive property for a given similarity measure. Fig.~\ref{fig:siamese1} shows the structure of the proposed model that combines two identical neural networks sharing the same structure and the same parameters $\Theta$. As input, we provide the model with the set $\mathcal{G}$ of all possible distinct pairs of stack traces encoded as an ordered sequence of stack frames. Each distinct frame is then referred to as a feature. The model output is provided with the similarity values for each pair of stack traces.  
The model $F$, with its corresponding parameters $\Theta$, consists in encoding a stack trace into a compact vector that represents a family of $M$ concatenated binary hash codes, each of which is encoded in $b$ bits in the Hamming space, denoted as $F_\Theta(s)$. The model consists in a concatenation of stacked convolution layers with different kernel sizes. We depict three kernel region sizes: $2, 3$ and $4$, each of which has $256$, $512$ or $1024$ filters. These filters perform convolutions on the one-hot encoded stack frames to generate feature maps. Then, $1$-Max pooling is performed over each map to record the largest number from each feature. Finally, the resulting features are concatenated to form a feature encoding vector for the penultimate layer that is fully connected to the hash model. It is noteworthy that any feature encoder structure (e.g., CNN, CNN-LSTM, AE, etc.) can be also used as a stack trace encoder instead of our proposed network architecture.  

Given the two hash vectors of the Siamese neural network, our contribution consists in designing an objective loss function that efficiently conducts $F_\Theta$ to learn a family of binary hash functions that aim to converge to the locality-sensitive property for the given similarity function $sim$. We propose to leverage Property~\eqref{eq:collisionProb} which is sufficient to imply the two required conditions of an LSH family. Assuming that the function $g$ is the identity function: $g(x) = I_x$ since the similarity values are within the closed interval $[0,1]$, the set of parameters $\Theta$ are optimized such that the probability of two random projected hash functions of order $k$ from the resulting hash vectors, $h_k^{i}$ and $h_k^{j}$ being equal, converges to the similarity value between the two stack traces $s_i$ and $s_j$ i.e.,

{\small
\begin{equation}
Pr[h_k^{i} = h_k^{j}] = Pr[H(s_i)_k = H(s_j)_k] = sim(s_i, s_j)
\label{eq:probaSimSiamese}
\end{equation}}

More formally, we seek to minimize the Mean Squared Error (MSE) between the probability of collision of two randomly projected hash functions of order $k$, i.e. $h_k^i$ resp. $h_k^j$ of $H(s_i)$ resp. $H(s_j)$, and the similarity value $sim(s_i,s_j)$, that is:
{ 
\begin{equation}
\argmin_{\Theta} \sum_{\substack{(s_i, s_j) \in \mathcal{G}}} \frac{1}{|\mathcal{G}|} [Pr[F_\Theta(s_i)_k = F_\Theta(s_j)_k] - sim(s_i,s_j)]^2 
\label{eq:lossSiamese}
\end{equation}}

At this point, the challenge is to formalize the probability of collision in the loss function. In other words, we attempt to quantify the probability $Pr[F_\Theta(s_i)_k = F_\Theta(s_j)_k]$ during the learning phase. In the following, we present the complete procedure for designing a computed loss function for the model $F$.   

\subsection{Objective loss function}

Given that the hash vectors are in a Hamming space, i.e., the vectors are restricted to binary values, $\{0,1\}$ or $\{-1,1\}$, it can be demonstrated that calculating the collision probability between two randomly projected hash functions of the same order $h_k^i$ and $h_k^j$ is equivalent to computing the Hamming similarity between the two hash vectors $H(s_i)$ and $H(s_j)$. This equivalence is satisfied since, for the Hamming similarity when $b = 1$, it has been proven in \cite{DBLP:conf/stoc/IndykM98}, that the projection function (i.e., a single bit drawn randomly) verifies the locality-sensitive property. In other terms, for two binary vectors $x$ and $x'$ of length $d$ with a Hamming distance $r$, the collision probability by randomly pulling a hash function from the set $\{h:[-1,1]^d \to \{-1,1\} \mid h(x) = x_i, i \in \{1,...,d\}\}$ verifies:

{ 
\begin{equation}
    Pr[h(x)=h(x')] = g(r) = 1-\frac{r}{d}
\label{eq:collisionProbHamming}
\end{equation}}


Intending to leverage property~\eqref{eq:collisionProbHamming} and given that our hash functions are rather $b$-bit encoded ($b \geq 1$), i.e., not restricted to a single projection but a succession of $b$ coordinates, we need to generalize this property for $b \geq 1$. Consequently, we define a generalized Hamming distance between two hash vectors $H(s_i)$ and $H(s_j)$ as the number of different projected hash functions of order $k$: $|\{k \in [\![1,M]\!] \mid h_k^i \ne h_k^j\}|$. 
As a result, each hash function that belongs to $\{h':[-1,1]^{M \times b} \to \{-1,1\}^b \mid h'(s) = H(s)_k \}$ satisfies the locality-sensitive property. This leads to conclude that for two different stack traces $s_i$ and $s_j$, the collision probability between two projected hash functions of a specific order $k'$ referring to~\eqref{eq:collisionProbHamming}:
{ 
\begin{equation}
    Pr[h_{k'}^i=h_{k'}^j] = 1-\frac{|\{k \in [\![1,M]\!] \mid h_k^i \ne h_k^j\}|}{M}
\label{eq:collisionProbHammingCusto}
\end{equation}}

Correspondingly, referring to the property~\eqref{eq:collisionProbHammingCusto}, the objective function as described in~(\ref{eq:lossSiamese}) can be formalized as follows:  

{ 
\begin{equation}
\sum_{\substack{(s_i, s_j) \in \mathcal{G}}} \frac{1}{|\mathcal{G}|}[1-\frac{|\{k \mid h_k^i \ne h_k^j, k \in [\![1,M]\!]\}|}{M} - sim(s_i,s_j)]^2 
\label{eq:lossSiameseFormalized}
\end{equation}}

A challenging problem in hashing, on the other hand consists in dealing with the binary constraint on hash vectors. This binary constraint leads to NP-hard mixed integer optimization problem~\cite{DBLP:conf/eccv/DoDC16}. In particular, the challenge in neural network parameter optimization is the \textit{vanishing gradient descent} from the \texttt{Sign} function used to obtain binary values. Specifically, the gradient of the \texttt{Sign} function is zero for all non-zero input values, which is limiting for neural networks that rely on gradient descent for training. 
In order to handle this challenge, most deep hashing techniques relax the constraint during the learning of hash functions using \texttt{Sigmoid} or \texttt{Hyperbolic Tangent} functions~\cite{DBLP:conf/cvpr/Liu0SC16,DBLP:conf/cvpr/0003CBS18,DBLP:conf/cvpr/LaiPLY15,DBLP:journals/tip/HuFXSCM18}. With this relaxation, the continuous hash codes are learned first. Then, the codes are binarized with thresholding.  
Continuous relaxation is a simple approach to address the original binary constraint problem. However, with binary hash codes that result from thresholding in the test phase, the solution may be suboptimal, compared to including the binary constraint in the learning phase. 
 
To this extent, we propose a simple yet efficient solution to cope with the binary constraint in the training phase. The solution lies in using approximate Hamming similarity. It requires having continuous values that are extremely close to binary values $\{-1,1\}$. We propose to use the  \texttt{Hyperbolic Tangent} activation on the hash layer while including the following condition in the loss function to drive the absolute hash values to be exceedingly close to $1$:
{ 
\begin{equation}
\frac{1}{M \cdot b} H(s)^T. H(s) - 1 = 0.
\label{eq:regularization1}
\end{equation}}

Under this regularization term incorporated into the loss function, we define the approximate generalized Hamming similarity as follows:

{ 
\begin{equation}
gHam \left(H(s_i), H(s_j)\right) = 1 - \frac{\sum_{k = 1}^M D_{\text{Chebyshev}}(h_k^i ,h_k^j)}{2 \cdot M}, \\
\label{eq:Hammingcase1}
\end{equation}}

where $D_{\text{Chebyshev}} = \max_{l \in \{1,...,b\}}(|h_{k,l}^i - h_{k,l}^j|)$ 

\smallbreak
The Chebyshev distance between $h_{k,l}^i$ and $h_{k,l}^j$ is then given as the maximum absolute distance in one of the $b$ dimensions. This implies that two hash codes are assumed to be similar if all bits of the hash code are matched for a specific projection. In other words, if $\exists$ $l \in \{1,...,b\}$ for a specific $k$ such that  $|h_{k,l}^i - h_{k,l}^j| \approx 2$, then $h_k^i$ and $h_k^j$ are considered as two different hash codes.

Finally, to ensure independence between the hash code bits along with the load-balanced locality-sensitive hashing, and inspired by the work of~\cite{DBLP:conf/eccv/DoDC16}, we have introduced the following regularization term that pushes the model to diversify the hash codes:
{\small
\begin{equation}
\frac{1}{M \cdot b} H(s)^T. \mathds{1}_{M \cdot b} = 0.
\label{eq:regularization2}
\end{equation}}

\noindent \textbf{Putting all together.} Having all the necessary elements to design an appropriate objective loss function to be optimized for {\sc DeepLSH} model, we define for convenience the following notations. Let $S = \{ sim(s_i, s_j)\}_{i,j \in [\![1,N]\!]} \in [0,1]^{N \times N}$ be the matrix representation of the similarities between all the stack trace pairs, and $\mathcal{H} = [H(s_1), H(s_2), ..., H(s_N)] \in [-1,1]^{M \cdot b \times N}$ be the approximate binary hash vectors generated by the model $F_\Theta$, such that, $H(s_i) = [h_1^i, h_2^i, ..., h_M^i]^T \in [-1,1]^{M \cdot b}$. We refer to $\mathcal{W} = \{ gHam (H(s_i), H(s_j))\}_{i,j \in [\![1,N]\!]} \in [0,1]^{N \times N}$ as the matrix representation of the generalized Hamming similarity between all pairs of hash vectors produced from the model $F_\Theta$. We formulate the following optimization problem to learn the parameters of our {\sc DeepLSH} model using gradient descent as follows:
{\small
\begin{align}
\min_{\Theta } \mathcal{L}_{\text{{\sc DeepLSH}}} &= \frac{1}{|\mathcal{G}|} \lVert \mathcal{W} - S \rVert^2 \nonumber \\
&+  \frac{\lambda_1}{2} \lVert \frac{1}{M \cdot b} \mathcal{H}^T \mathcal{H} - \textbf{I}_{N} \rVert^2 \nonumber \\ 
&+ \frac{\lambda_2}{|\mathcal{G}|} \lVert \frac{1}{M \cdot b} \mathcal{H}^T \mathds{1}_{M \cdot b} \rVert^2 \nonumber \\
&+ \lambda_3 \lVert \Theta \rVert^2_F,
\label{eq:generalLoss}
\end{align}}

where $\lambda_1$, $\lambda_2$ and $\lambda_3$ are regularization parameters to assess the importance of the different parts of the objective function.

\section{Experiments}\label{sec:xp}

\subsection{Experimental Setup and Evaluation Protocol}
We report our experimental study to assess the effectiveness of {\sc DeepLSH} in performing efficient, fast, and scalable approximate nearest neighbors search, by providing appropriate hash functions that can approximate the stack trace similarity measures and allow them to scale when used in large databases of bug reports.

\smallbreak
\noindent \textbf{Stack trace dataset and training methodology.} Our experiments are conducted on a real-world dataset comprising stack traces automatically reported by our ERP software. To establish a robust training dataset, we selectively choose the most frequent stack traces from our historical incident database, creating distinct pairs of stack traces. These pairs are then utilized to train our {\sc DeepLSH} model. Each pair is assigned a similarity value calculated using diverse similarity functions. We evaluate the performance of {\sc DeepLSH} on twelve different similarity measures: Jaccard (bag-of-words and bi-grams), Cosine (bag-of-words, bi-grams, and TF-IDF), Edit distance~\cite{DBLP:conf/osdi/BartzSPKGCL08}, PDM~\cite{DBLP:conf/icse/DangWZZN12}, Brodie~\cite{DBLP:conf/icac/BrodieMLMMWCS05}, DURFEX~\cite{DBLP:conf/qrs/SaborHL17}, Lerch~\cite{DBLP:conf/csmr/LerchM13}, Moroo~\cite{DBLP:conf/seke/MorooAH17}, and TraceSim~\cite{DBLP:conf/sigsoft/VasilievKCKLP20}. It is important to note that these similarity metrics serve as a reference point and act as the ground truth in our current setup. Our objective, therefore, is not to evaluate the effectiveness of these similarity measures or compare them with each other. This is because each measure is applied to a vast dataset of stack traces, where labeled information is not always available, or manual labeling is impractical, especially in cases of frequent background process failures that may occur in the thousands per day. Our goal is to ensure that regardless of the used similarity measure employed for stack traces, {\sc DeepLSH} approach can effectively replicate this measure of similarity. Additionally, it should be capable of scaling up and being utilized within large-scale systems.

\begin{figure*}
\centering
 \includegraphics[width=0.95\textwidth]{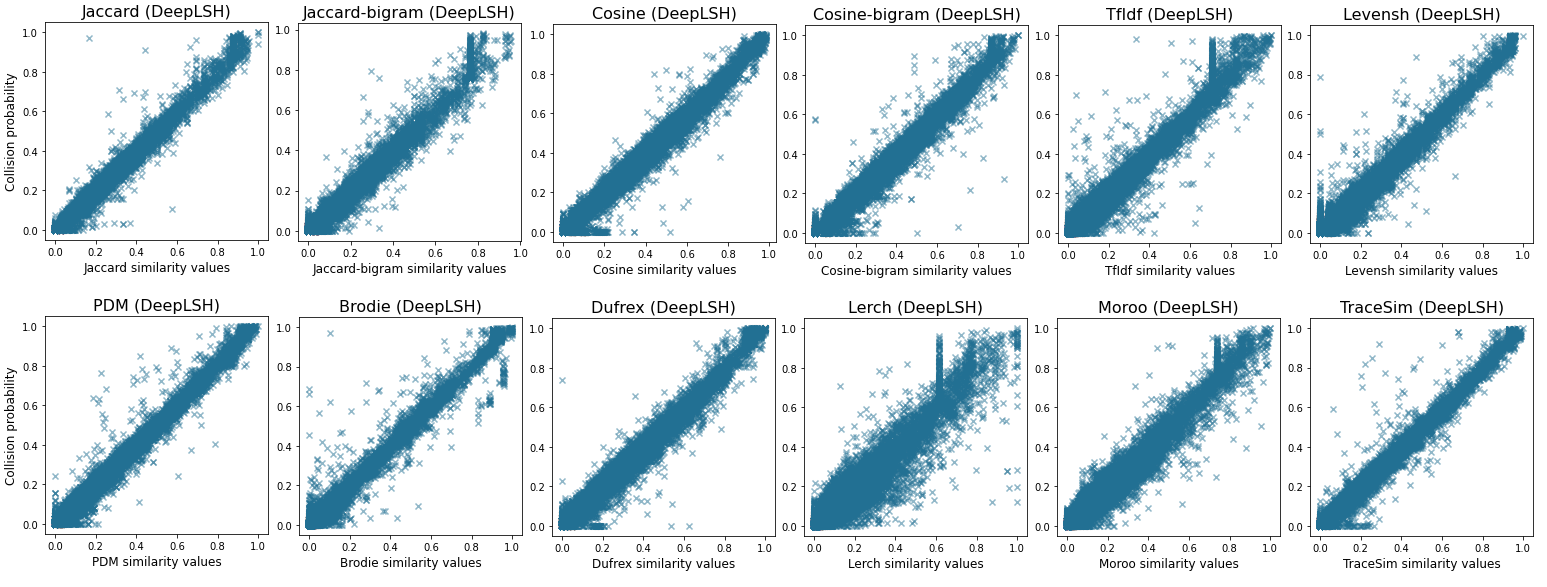}
\caption{\label{fig:corrLSH}   Locality-sensitive preserving: Correlation between the probability of hash collision and the similarity value}
\end{figure*}

Regarding the training methodology, the training set consists of $499500$ pairs of stack traces, while the validation and test set are constituted of $99900$ pairs. The number of hash functions $M$ and the size of each hash code $b$ can be parameterized by the user. By default these values are respectively set to $64$ hash functions of $8$ bits. The max iteration is fixed at $20$ epochs with a batch size of $256$ or $512$. The parameter optimization process is achieved with the readily available \texttt{Adam} optimizer of \texttt{TensorFlow} with an adaptive learning rate and a weight decay of $1e^{-4}$. With this configuration, the training process takes barely $10$ minutes. The source code and data with all associated instructions required for experimental replication are made available~\footnote{\url{https://github.com/RemilYoucef/deep-locality-sensitive-hashing}}

\begin{figure}
\centering
 \includegraphics[width=0.45\textwidth]{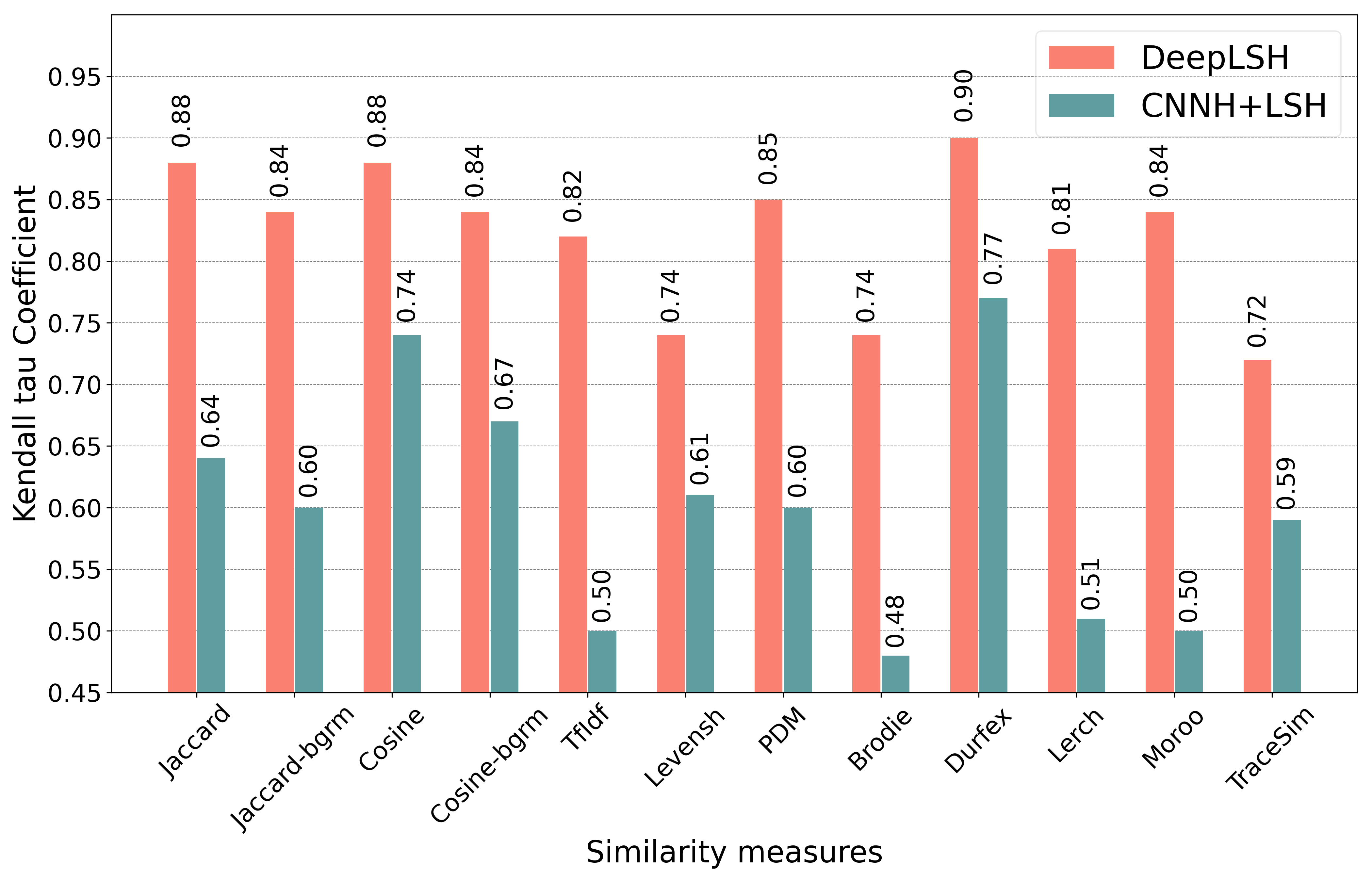}
\caption{\label{fig:corrCoefs} Kendall's $\tau$ coefficient between the real and predicted pairwise similarities}
\end{figure}

\smallbreak 
\noindent \textbf{Baselines.} As there is no explicit competing approach for {\sc DeepLSH} in the state-of-the-art, we chose to initially compare with \textbf{(1) Standard LSH methods}, namely Min-Hash and Sim-Hash. This comparison should only be performed with the Jaccard and Cosine metrics respectively since they are not generalizable to other measures compared to {\sc DeepLSH}. As mentioned in Sec~\ref{sec:relatedwork}, we compare against the closest method to our work referred to as \textbf{(2)} \texttt{CNNH+LSH}~\cite{DBLP:journals/tip/HuFXSCM18}. This approach uses the concept of learn to hash and then performs in a post-processing step the Hamming LSH. The methodology followed in this work is significantly different to ours. {\sc DeepLSH} unlike the latter incorporates the LSH component into the model learning phase, resulting in a new loss function with related regularization to meet the locality-sensitive property. Regarding the application of~\cite{DBLP:journals/tip/HuFXSCM18} on stack-traces, since it was primarily designed for images, we only needed to provide a list of one-hot encoded stack-frames to the convolutional feature extractor instead of pixels. Finally, to evaluate the scalability of our approach, we compare against \textbf{(3) Native k-NN (k-Nearest neighbors)} approach of linear complexity, using the exact computation of similarity functions between stack traces. It is noteworthy that clustering techniques have not been considered as baselines (as discussed in the introduction), since the addressed problem is an ANN search.

\smallbreak
\noindent \textbf{Evaluation protocol.} Through this experimental study, we address the following research questions by proposing an evaluation protocol to assess each point claimed in this work: \\
\textbf{RQ1 [Model Evaluation]}: Does {\sc DeepLSH} model manage to converge to the locality-sensitive property to mimic a diverse set of stack-trace-based similarity metrics? We first highlight, by means of the Kendall $\tau$ ranking coefficient~\cite{kendall1938new}, whether the model succeeds in preserving the original order between the predicted pairwise similarities. In addition, we study how accurately the generalized Hamming similarity approximates the true Hamming similarity between the discretized hash vectors in the test phase. \\
\textbf{RQ2 [{\sc DeepLSH} for ANN search]}: Does {\sc DeepLSH} model achieve satisfactory performance in finding near-duplicate crash reports using a given similarity measure? By querying the model to obtain the hash vectors for unseen stack traces, we study the search performance to retrieve approximately near-duplicate stack traces based on two metrics that are widely used in the context of crash deduplication: Recall Rate at the first $k$ positions (RR@$k$)~\cite{DBLP:journals/ftir/Sanderson10} and the Mean Reciprocal Rank (MRR)~\cite{Craswell2009}. \\
\textbf{RQ3 [Preserving LSH guarantees]}: To what extent does {\sc DeepLSH} succeed in preserving the guarantees of LSH compared to Standard LSH methods and the baseline (\texttt{CNNH+LSH})~\cite{DBLP:journals/tip/HuFXSCM18}. For this purpose, we provide recall, precision and F-score measures adjusted to quantify the extent to which the probability-similarity constraint~\eqref{eq:relationProbSim} has been satisfied (more details are provided hereafter). \\ 
\textbf{RQ4 [Runtime Analysis]}: How does the scalability of DeepLSH compare to that of a native linear k-NN approach? We report the execution time required for {\sc DeepLSH} to find the near-duplicates, compared to a k-NN approach of linear time complexity.

\subsection{RQ1: Model Evaluation}

\begin{figure}
\centering
 \includegraphics[width=0.45\textwidth]{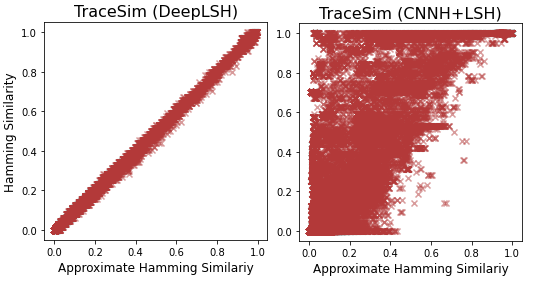}
\caption{\label{fig:corrHamm}   Comparison between {\sc DeepLSH} and~\cite{DBLP:journals/tip/HuFXSCM18} in preserving the Hamming similarity between hash vectors}
\end{figure}

In Fig.~\ref{fig:corrLSH}, we highlight the strong linear correlation between the probability of hash collision and the similarity values for almost all similarity measures performed on stack trace pairs. The resulting plots show that the model is able to converge perfectly to the locality-sensitive property for almost all similarity measures. We observed a few outliers in the TF-IDF, Lerch, and Moroo measures. These outliers can be attributed to extremely low or high IDF values, indicating frames that are either non-discriminatory or infrequent among stack traces. Fortunately, their presence does not significantly affect the model's performance. However, to further address this issue, we can consider augmenting the feature set by including the IDF of the frame features from the training set. It is also important to assess the capability of {\sc DeepLSH} model to maintain the order between pairwise similarity values. For instance, for a triplet of stack traces $s, p$ and $q$ if $sim(s,p) > sim(s, q)$, we aim to evaluate whether the model is likely to provide hash functions s.t. $Pr[h_k(s) = h_k(p)]  > Pr[h_k(s) = h_k(q)]$ for $k \in M$. For this purpose, we measure the Kendall rank correlation coefficient, between the set of similarity values, and the set of generalized Hamming similarities between the resulting hash vectors as shown in Fig.~\ref{fig:corrCoefs}. Remarkably, we obtained satisfactory results compared to our baseline (on average, $0.82$ for {\sc DeepLSH}, against $0.60$ for \texttt{CNNH+LSH}, that is, $0.22$ of improvement) which permitted to achieve better and accurate results on the ANN search. Finally, thanks to the regularization conditions~\eqref{eq:regularization1} and~\eqref{eq:regularization2} incorporated in the objective loss function, the model yields approximate binary hash values extremely close to $\{-1,1\}$ that are binarized/relaxed in the test phase. We seek to evaluate whether our proposed solution to deal with the binarization problem using the generalized Hamming similarity~\eqref{eq:Hammingcase1} performed in the training phase, is optimal and captures the true Hamming similarity between the discritized hash values in the test phase. Considering the TraceSim measure as an example in Fig.~\ref{fig:corrHamm} (on the left), we notice a strong linear correlation between the true hamming similarity calculated on the binary vectors and the approximate generalized Hamming similarity used in the loss function during the learning phase. This means that our loss optimization process is as identical as the optimization of any loss function with strictly binary values in the training phase. In the same Figure (on the right), we show the impact of not incorporating the LSH component into the model, as has been done in~\cite{DBLP:journals/tip/HuFXSCM18}. Performing LSH on the discretized vectors in a post-processing step results in a sub-optimal optimisation, since the correlation between the similarity calculated using basic embedding and the true Hamming similarity is not even monotonic. Consequently, \texttt{CNNH+LSH} has failed to capture TraceSim similarity.

\subsection{RQ2: Evaluation of {\sc DeepLSH} for ANN}

\begin{table*}
    \centering
    \caption{  Comparison between the search performances of {\sc DeepLSH} against the standard LSH approaches w.r.t. their addressed similarity measures and (\texttt{CNNH+LSH})~\cite{DBLP:journals/tip/HuFXSCM18} in terms of Recall Rate (RR@$k$) and Mean Reciprocal Rank (MRR).}
    \renewcommand{\arraystretch}{1}
    \resizebox{\textwidth}{!}{\begin{tabular}{ccccccccccccc}
    \toprule
    \multirow{2}{*}{\textbf{Similarity Measure}} & \multicolumn{4}{c}{\textbf{RR@$1$}}   & \multicolumn{4}{c}{\textbf{RR@$5$}}     & \multicolumn{4}{c}{\textbf{MRR}}     \\* \cmidrule(l){2-13} 
                                        & \textcolor{red}{\texttt{CNNH+LSH}} & \textcolor{PineGreen}{{\sc DeepLSH}} & \textcolor{blue}{MinHash}   & \textcolor{blue}{SimHash} & \textcolor{red}{\texttt{CNNH+LSH}} & \textcolor{PineGreen}{{\sc DeepLSH}} & \textcolor{blue}{MinHash}   & \textcolor{blue}{SimHash} & \textcolor{red}{\texttt{CNNH+LSH}} & \textcolor{PineGreen}{{\sc DeepLSH}} & \textcolor{blue}{MinHash}   & \textcolor{blue}{SimHash}    \\* \midrule
    Jaccard                             & $0.71$   & $0.87$  & $\boldsymbol{0.90}$ & $-$   & $0.79$  & $\boldsymbol{0.92}$ & $\boldsymbol{0.92}$   & $-$  & $0.85$ & $\boldsymbol{0.96}$   & $0.95$  & $-$ \\
    Jaccard-bigram                      & $0.67$   & $0.87$  & $\boldsymbol{0.90}$ & $-$   & $0.76$  & $0.91$ & $\boldsymbol{0.92}$   & $-$  & $0.82$ & $0.93$   & $\boldsymbol{0.94}$  & $-$ \\
    Cosine                              & $\boldsymbol{0.84}$   & $0.81$  & $-$  & $0.61$   & $0.83$  & $\boldsymbol{0.90}$ & $-$   & $0.62$  & $\boldsymbol{0.88}$ & $0.87$   & $-$  & $0.80$ \\
    Cosine-bigram                       & $0.76$   & $\boldsymbol{0.84}$  & $-$  & $0.58$   & $0.79$  & $\boldsymbol{0.93}$ & $-$   & $0.58$  & $0.89$ & $\boldsymbol{0.91}$   & $-$  & $0.80$ \\
    TF-IDF                              & $0.73$   & $\boldsymbol{0.76}$  & $-$  & $0.55$   & $0.75$  & $\boldsymbol{0.88}$ & $-$   & $0.55$  & $0.85$ & $\boldsymbol{0.90}$   & $-$  & $0.73$ \\
    Edit Distance~\cite{DBLP:conf/osdi/BartzSPKGCL08}                       & $0.81$   & $\boldsymbol{0.88}$  & $-$  & $-$   & $0.75$  & $\boldsymbol{0.94}$ & $-$   & $-$  & $0.88$ & $\boldsymbol{0.95}$   & $-$  & $-$ \\
    PDM~\cite{DBLP:conf/icse/DangWZZN12}                                & $0.80$   & $\boldsymbol{0.84}$  & $-$  & $-$   & $0.76$  & $\boldsymbol{0.90}$ & $-$   & $-$  & $0.82$ & $\boldsymbol{0.93}$   & $-$  & $-$ \\
    Brodie~\cite{DBLP:conf/icac/BrodieMLMMWCS05}                        & $0.79$   & $\boldsymbol{0.84}$  & $-$  & $-$   & $0.76$  & $\boldsymbol{0.90}$ & $-$   & $-$  & $0.82$ & $\boldsymbol{0.93}$   & $-$  & $-$ \\
    DURFEX~\cite{DBLP:conf/qrs/SaborHL17}                               & $0.72$   & $\boldsymbol{0.83}$  & $-$  & $-$   & $0.79$  & $\boldsymbol{0.91}$ & $-$   & $-$  & $0.82$ & $\boldsymbol{0.91}$   & $-$  & $-$ \\
    Lerch~\cite{DBLP:conf/csmr/LerchM13}                    & $0.70$   & $\boldsymbol{0.78}$  & $-$  & $-$   & $0.70$  & $\boldsymbol{0.85}$ & $-$   & $-$  & $0.80$ & $\boldsymbol{0.88}$   & $-$  & $-$ \\
    Moroo~\cite{DBLP:conf/seke/MorooAH17}                         & $0.75$   & $\boldsymbol{0.80}$  & $-$  & $-$   & $0.68$  & $\boldsymbol{0.90}$ & $-$   & $-$  & $0.80$ & $\boldsymbol{0.93}$   & $-$  & $-$ \\
    TraceSim~\cite{DBLP:conf/sigsoft/VasilievKCKLP20}                            & $\boldsymbol{0.81}$   & $0.79$  & $-$  & $-$   & $0.75$  & $\boldsymbol{0.90}$ & $-$   & $-$  & $0.84$ & $\boldsymbol{0.92}$   & $-$  & $-$ \\* \bottomrule
    
    \end{tabular}}
  \label{tab:searchPerfs}
\end{table*}

The objective of our {\sc DeepLSH} approach, given a similarity measure, is to generate, for a new stack trace $s$ reported by our monitoring system, an appropriate hash vector to query a $(L,K)-$parameterized LSH for quickly and efficiently locate in a sub-linear time complexity its near-duplicates. The hash vector contains $M$ hash functions, partitioned across $L$ hash tables, each consisting of a concatenation of $K$ hash functions called a meta-hash function of size $K \cdot b$. A stack trace $q \in \mathcal{R}_s$ is identified as a near-duplicate stack of $s$ if it matches the stack trace $s$ at least in one meta-hash function. The set of near-duplicate stack traces of $s$ is denoted by $\mathcal{R}_s$. It is worth noting that the set $\mathcal{R}_s$ is sorted in the original order with respect to the similarity measure value. 

\begin{figure}
\centering
 \includegraphics[width=0.48\textwidth]{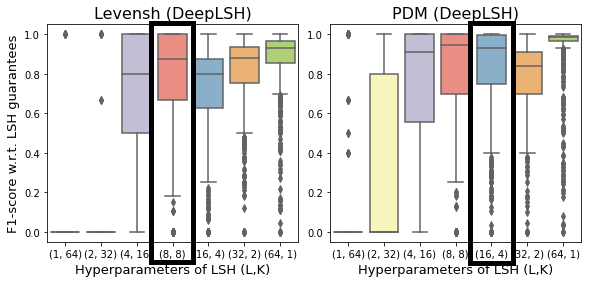}
\caption{\label{fig:fscore}   F-score boxplots w.r.t. different values of (L,K)}
\end{figure}

LSH offers the possibility to control the trade-off between precision and recall (w.r.t LSH guarantees) by setting the hyperparameters values $K$ and $L$. We can simply choose to consider the tuple of hyperparameters that maximizes the F-score. However, as shown in Fig.~\ref{fig:fscore} (e.g., $(L,K) = (16,4)$ for PDM), we ignore extreme cases i.e., cases where the threshold is very small or very large, which refers to a very large value of $L$ or $K$, or cases where the variance is very high (for more details on how the F-score is calculated, refer to~\ref{subsec:lshGuarantees}). As an example, when using the combination $(L,K) = (64,1)$, we observe higher F-score values. However, it's important to note that this is partly a result of selecting a very low threshold, which in turn leads to a larger number of near-duplicates that need to be analyzed.

In a first analysis, we were interested in the \textit{recall rate of order} $k$. For each stack trace $s$ that belongs to a query set $\mathcal{Q}$, we yield a set of its approximate nearest neighbors $\mathcal{R}_s$ (i.e., potential near-duplicates) such that $|\mathcal{R}_s| \geq k$ and hence,     
$$ 
\textbf{\text{RR}@k} = \frac{1}{k \cdot |\mathcal{Q}|} \sum_{s \in \mathcal{Q}} \sum_{i = 1}^k  \mathds{1}_{[nn_i(s, \text{*args}) \in \mathcal{R}_s]},
$$
where $nn_i(s, \text{*args})$ is a function that returns the real nearest neighbor of order $i$ for the stack trace $s$ given a set of historical stack traces and the LSH hyperparameters $L$ and $K$.

In order to evaluate the ranking quality of a set of near-duplicates $\mathcal{R}_s$ for a stack trace $s$ according to a $(L,K)$ combination, and relative to the set of true nearest neighbors $\mathcal{T}_s$, we use the \textit{Mean reciprocal rank} (MRR)~\cite{Craswell2009}. This measure seeks to compute the reciprocal rank of a retrieved near-duplicate $s' \in \mathcal{R}_s$ relative to its actual position in the set of true nearest neighbors. More concretely:
$$
\textbf{\text{MRR}} = \frac{1}{|\mathcal{Q}|} \sum_{s \in \mathcal{Q}} \frac{1}{|\mathcal{R}_s|} \sum_{s' \in \mathcal{R}_s} \frac{rank(s',\mathcal{R}_s)}{rank(s',\mathcal{T}_s)}
$$
E.g., let's consider a given stack trace $q$, where we retrieve the set of its approximate nearest neighbors and subsequently sort them according to the original order $\mathcal{R}_q = \{ s_2, s_3, s_5\}$ and the set of its true nearest neighbor is given as $\mathcal{T}_q = \{ s_1, s_2, s_3, s_4, s_5\}$. The MRR is then: $\frac{1}{3}(\frac{1}{2} + \frac{2}{3} + \frac{3}{5}) \approx 0.63$. The MRR in this case is over penalized since we failed to find the true nearest neighbor $s_1$ in $\mathcal{R}_q$. It is noteworthy that the set $\mathcal{R}_q$ is sorted according to the original order w.r.t to the similarity measure.   

\begin{table*}
    \centering
    \caption{  Comparison between the precision/recall and f-score of {\sc DeepLSH} in preserving the probability-similarity relation~\eqref{eq:relationProbSim} against the standard LSH approaches w.r.t. theirs addressed similarity measures and (\texttt{CNNH+LSH})~\cite{DBLP:journals/tip/HuFXSCM18}.}
    \renewcommand{\arraystretch}{1}
    \resizebox{\textwidth}{!}{\begin{tabular}{ccccccccccccc}
    \toprule
    \multirow{2}{*}{\textbf{Similarity Measure}} & \multicolumn{4}{c}{\textbf{Precision}}   & \multicolumn{4}{c}{\textbf{Recall}}     & \multicolumn{4}{c}{\textbf{F-score}}     \\* \cmidrule(l){2-13} 
                                        & \textcolor{red}{\texttt{CNNH+LSH}} & \textcolor{PineGreen}{{\sc DeepLSH}} & \textcolor{blue}{MinHash}   & \textcolor{blue}{SimHash} & \textcolor{red}{\texttt{CNNH+LSH}} & \textcolor{PineGreen}{{\sc DeepLSH}} & \textcolor{blue}{MinHash}   & \textcolor{blue}{SimHash} & \textcolor{red}{\texttt{CNNH+LSH}} & \textcolor{PineGreen}{{\sc DeepLSH}} & \textcolor{blue}{MinHash}   & \textcolor{blue}{SimHash}    \\* \midrule
    Jaccard                             & $0.64$   & $\boldsymbol{0.78}$  & $0.76$ & $-$   & $0.78$  & $\boldsymbol{0.85}$ & $\boldsymbol{0.85}$   & $-$  & $0.70$ & $\boldsymbol{0.81}$   & $0.80$  & $-$ \\
    Jaccard-bigram                      & $0.56$   & $\boldsymbol{0.76}$  & $0.70$ & $-$   & $0.70$  & $0.74$ & $\boldsymbol{0.83}$   & $-$  & $0.62$ & $\boldsymbol{0.75}$   & $\boldsymbol{0.75}$  & $-$ \\
    Cosine                              & $\boldsymbol{0.77}$   & $0.72$  & $-$  & $0.74$   & $0.74$  & $\boldsymbol{0.84}$ & $-$   & $0.6$  & $0.75$ & $\boldsymbol{0.78}$   & $-$  & $0.66$ \\
    Cosine-bigram                       & $\boldsymbol{0.74}$   & $\boldsymbol{0.74}$  & $-$  & $0.66$   & $0.72$  & $\boldsymbol{0.82}$ & $-$   & $0.41$  & $0.73$ & $\boldsymbol{0.78}$   & $-$  & $0.50$ \\
    TF-IDF                              & $\boldsymbol{0.85}$   & $0.76$  & $-$  & $0.49$   & $0.61$  & $\boldsymbol{0.86}$ & $-$   & $0.62$  & $0.71$ & $\boldsymbol{0.81}$   & $-$  & $0.55$ \\
    Edit Distance~\cite{DBLP:conf/osdi/BartzSPKGCL08}                       & $0.37$   & $\boldsymbol{0.78}$  & $-$  & $-$   & $0.78$  & $\boldsymbol{0.88}$ & $-$   & $-$  & $0.50$ & $\boldsymbol{0.83}$   & $-$  & $-$ \\
    PDM~\cite{DBLP:conf/icse/DangWZZN12}                                & $0.68$   & $\boldsymbol{0.85}$  & $-$  & $-$   & $0.76$  & $\boldsymbol{0.86}$ & $-$   & $-$  & $0.72$ & $\boldsymbol{0.85}$   & $-$  & $-$ \\
    Brodie~\cite{DBLP:conf/icac/BrodieMLMMWCS05}                        & $0.36$   & $\boldsymbol{0.83}$  & $-$  & $-$   & $0.81$  & $\boldsymbol{0.86}$ & $-$   & $-$  & $0.50$ & $\boldsymbol{0.84}$   & $-$  & $-$ \\
    DURFEX~\cite{DBLP:conf/qrs/SaborHL17}                               & $0.73$   & $\boldsymbol{0.78}$  & $-$  & $-$   & $0.70$  & $\boldsymbol{0.79}$ & $-$   & $-$  & $0.71$ & $\boldsymbol{0.78}$   & $-$  & $-$ \\
    Lerch~\cite{DBLP:conf/csmr/LerchM13}                    & $0.74$   & $\boldsymbol{0.76}$  & $-$  & $-$   & $0.57$  & $\boldsymbol{0.76}$ & $-$   & $-$  & $0.64$ & $\boldsymbol{0.76}$   & $-$  & $-$ \\
    Moroo~\cite{DBLP:conf/seke/MorooAH17}                         & $0.66$   & $\boldsymbol{0.73}$  & $-$  & $-$   & $\boldsymbol{0.85}$  & $0.82$ & $-$   & $-$  & $0.74$ & $\boldsymbol{0.77}$   & $-$  & $-$ \\
    TraceSim~\cite{DBLP:conf/sigsoft/VasilievKCKLP20}                            & $0.31$   & $\boldsymbol{0.80}$  & $-$  & $-$   & $0.84$  & $\boldsymbol{0.88}$ & $-$   & $-$  & $0.45$ & $\boldsymbol{0.84}$   & $-$  & $-$ \\* \bottomrule
    
    \end{tabular}}
  \label{tab:lshGuarantees}
\end{table*}

The results are presented in detail in Table~\ref{tab:searchPerfs}, according to the identified similarity measures that have been proposed to address the crash-deduplication problem. We choose 2 different values of $k = \{1,5\}$ for the recall rate. We compare {\sc DeepLSH} against MinHash, SimHash and \texttt{CNNH+LSH}. Ideally, standard LSH techniques should guarantee optimal search accuracy compared to {\sc DeepLSH}, since they are proven to converge to the locality-sensitive property w.r.t. their similarity measures. Interestingly, we observe that {\sc DeepLSH} almost matches the search performances of MinHash on Jaccard similarity, and outperforms SimHash. In fact, while MinHash is a reliable probabilistic model designed for estimating Jaccard similarity with guarantees, it lacks the versatility to be extended to other similarity metrics, especially those intended for comparing stack traces. This limitation significantly restricts its applicability to the crash deduplication problem. Jaccard measure, which it relies on, may not be the most suitable metric for stack trace comparison as it does not account for the order of frames or the sequential aspect of function invocation within stack traces. This demonstrates also that {\sc DeepLSH} is not only generalizable to other complex measures, but can even be used for measures where an existing LSH is already known. We also show that {\sc DeepLSH} outperforms \texttt{CNNH+LSH} with a large margin on $3$ different comparison metrics and for almost all similarity measures. More specifically, we notice that {\sc DeepLSH} search performance is enhanced with a larger value of $k$ up to $0.94$ for Edit distance with an improvement of $\sim 0.2$ over \texttt{CNNH+LSH}.

\subsection{RQ3: LSH guarantees Preserving}\label{subsec:lshGuarantees}

In what follows, we aim to evaluate whether {\sc DeepLSH} succeeds in preserving the guarantees of LSH regarding the probability-similarity relation in~\eqref{eq:relationProbSim}. To this end, for a specific stack trace $s$ we look for the true near-duplicate stack traces $q \in \mathcal{R}_s^{\text{True}}$ that the model should return with a $(K,L)$ setting, s.t. $P_{K,L}(s,q) = 1-(1-sim(s,q)^K)^L \geq 0.5$, i.e., the probability to belong to the set is equal or higher than $0.5$. We then derive the precision, recall and F-score between the returned set of near-duplicates $\mathcal{R}_s$ and $\mathcal{R}_s^{\text{True}}$. More formally we derive this values, s.t:  
$$\textbf{\text{Recall}} = \frac{1}{|\mathcal{Q}|} \sum_{s \in \mathcal{Q}} \frac{|\mathcal{R}_s \cap \mathcal{R}_s^{\text{True}}|}{|\mathcal{R}_s^{\text{True}}|}$$ 

$$\textbf{\text{Precision}} = \frac{1}{|\mathcal{Q}|} \sum_{s \in \mathcal{Q}} \frac{|\mathcal{R}_s \cap \mathcal{R}_s^{\text{True}}|}{|\mathcal{R}_s|}$$

$$\textbf{\text{F-score}} = \frac{2}{\text{Recall}^{-1}+\text{Precision}^{-1}}$$

\smallbreak

As detailed in Table~\ref{tab:lshGuarantees}, one can notice that {\sc DeepLSH} is much better than all baselines in terms of F-score on almost all similarity measures including the accurate Minhash for Jaccard. {\sc DeepLSH} showed significantly better performance in terms of recall, i.e., its ability to capture all similarities that are beyond the threshold imposed by an optimal parameterization of $(K,L)$. On the other hand, the reported precision values, as opposed to \texttt{CNNH+LSH}, show that {\sc DeepLSH} does not generate false positives, so that false near-duplicates are not grouped in the same bucket. 
In particular, on similarity measures that use the Levenshtein distance (e.g. ED, Brodie and TraceSim), we observe a rather low precision for \texttt{CNNH+LSH}, which shows on the one hand the limitation of \texttt{CNNH+LSH} to generalize to such metrics, and on the other hand agrees with the explanation of Fig.~\ref{fig:corrHamm}.

\begin{table}
    \centering
    \caption{  Comparison between the runtime required to find near-duplicate stack traces for {\sc DeepLSH}, k-NN based approach and standard LSH techniques.}
    \renewcommand{\arraystretch}{1}
    \scalebox{0.84}{\begin{tabular}{ccccccccccccc}
    \toprule
    \multirow{2}{*}{Similarity Measure} & \multicolumn{5}{c}{Runtime ($\sim$ Seconds)}        \\* \cmidrule(l){2-5} 
                                        & k-NN   &  \texttt{CNNH+LSH} & {\sc DeepLSH} & MinHash & SimHash \\* \midrule
    Jaccard                             & $258$  &   $30$     & $26$    &   $57$ &    -               \\
    Cosine                              & $8288$  &   $15$    & $14$    & -  & $3$               \\
    TF-IDF                              & $8510$   &   $16$   & $15$    &  - & $4$              \\
    Edit Distance~\cite{DBLP:conf/osdi/BartzSPKGCL08}                      & $4911$ &   $29$     & $29$    & - &  -                \\
    PDM~\cite{DBLP:conf/icse/DangWZZN12}                                 & $10047$  &  $16$    & $16$    & - & -                \\
    Brodie~\cite{DBLP:conf/icac/BrodieMLMMWCS05}                       & $\text{Limit}$& $27$ & $27$    &-  & -               \\
    DURFEX~\cite{DBLP:conf/qrs/SaborHL17}                              & $12160$  &   $26$   & $24$    &  - & -               \\
    Lerch~\cite{DBLP:conf/csmr/LerchM13}                    & $3118$  &   $24$    & $24$   & - &  -               \\
    Moroo~\cite{DBLP:conf/seke/MorooAH17}                        & $15253$ &   $25$    & $25$    & - &   -               \\ 
    TraceSim~\cite{DBLP:conf/sigsoft/VasilievKCKLP20}                            & $13050$  &  $30$         & $30$ & - & -\\* \bottomrule
    
    \end{tabular}}
  \label{tab:runtime}
\end{table}

\subsection{RQ4: Runtime Analysis}

We evaluate the scalability of {\sc DeepLSH} and how quickly this approach identifies, for a batch of stack traces in a large historical crash database, the most similar stacks w.r.t. a given similarity measure. We report the execution time required to find the near-duplicate candidates for $1,000$ new stack traces when querying on more than $100,000$ historical crashes (100 million queries). We compare basically against the native k-NN based approach for all similarity measures. Recall that uni-gram and big-gram representations yield identical results for the cosine, Jaccard, and TF-IDF measures. The reported results are depicted in Table~\ref{tab:runtime}. The execution time for a native k-NN approach depends on the batch size, the size of the database and the computational complexity of the similarity measure. With the latter, most similarity measures under the conditions described above require more than an hour to return any results (more than $3$ hours for DURFEX, Moroo and TraceSim) and no results returned by Brodie within 10 hours. On the other hand, {\sc DeepLSH} only depends on the number of hash tables which does not exceed $64$ tables. We notice that the runtime is roughly constant and around $24 \sim 27$ seconds on average. Remarkably, {\sc DeepLSH} is even faster than MinHash for Jaccard Similarity. SimHash, on the other hand, has proven to be faster, and \texttt{CNNH+LSH} has been similar to {\sc DeepLSH} in runtime, but as seen above, both approaches show poor search performance and lower guarantees.

\subsection{Experiments Discussion and Perspectives}

Through our experiments, we have successfully demonstrated the effective utilization of {\sc DeepLSH} w.r.t. a diverse range of similarity measures, specifically designed for comparing bug reports. By preserving the locality-sensitive hashing property, our approach offers reliable guarantees on the search accuracy of retrieving near-duplicates while significantly enhancing the retrieval time. When evaluated against baseline methods, {\sc DeepLSH}  exhibited highly satisfactory performance, successfully identifying the most similar stack traces based on the chosen similarity measure used as a reference point. Notably, {\sc DeepLSH} is a groundbreaking similarity-agnostic approach that applies hashing techniques in this specific context, while matching well-known hashing technique such as MinHash and even surpassing SimHash which are tailored for only a single similarity measure. It may be questioned why we don't simply employ MinHash with the Jaccard coefficient, as it already delivers satisfactory performance and scalable results. However, the Jaccard coefficient is not the most appropriate measure for comparing stack traces. That is why various similarity measures have been proposed in the field of crash deduplication to address the unique characteristics of stack traces. Unfortunately, MinHash is not suitable for these alternative similarity measures and cannot be effectively applied. Furthermore, our runtime analysis has revealed that {\sc DeepLSH} exhibits impressive computational efficiency, regardless of the complexity of the chosen similarity measure. This remarkable finding allows users to freely choose any similarity measure suitable for their system, ensuring both high search accuracy and fast retrieval. Moreover, our method does not only offer users the flexibility to adjust the trade-off between recall and precision through the values of $K$ and $L$ but also provides a systematic approach to selecting the optimal hyperparameters' combination that maximizes the F-score w.r.t. LSH property preserving, as explained in Figure~\ref{fig:fscore}. It's worth mentioning that augmenting the number of hash tables may slightly increase computation time but remains well within the order of a few seconds.  Overall, the experimental results clearly demonstrate the superiority of {\sc DeepLSH} in terms of both performance and flexibility and offer an innovative solution for fast near-duplicate stack traces retrieval.

As a future perspective, we intend to evaluate our approach on other large public datasets of stack traces, which might be available but not exclusively specified for Java stack traces. It is also noteworthy that this method can be easily adapted to various settings, applied to other datasets, and generalized to many other use cases. For instance, it could be utilized for text similarity tasks with complex metrics like Word Mover's Distance~\cite{DBLP:conf/nips/HuangGKSSW16}, in conjunction with natural language processing techniques. This area presents an exciting opportunity for exploration that we plan to delve into.

\section{Conclusion \label{ref:conclusion}}

In this paper, we tackle the important task of fast and efficient automatic crash bucketing in software development. We investigate the potential of locality-sensitive hashing (LSH) for this purpose, leveraging its sublinear performance and theoretical guarantees in terms of accuracy for similarity search. This approach offers significant advantages when dealing with large datasets, yet surprisingly, LSH has not been explored in the crash bucketing literature. The main reason for the lack of consideration of LSH in crash bucketing research is the challenge of deriving hash functions that satisfy the locality-sensitive property for advanced and complex crash bucketing metrics. To address this gap, we propose a novel, parameterizable approach named {\sc DeepLSH}. We introduce an original objective loss function, complemented by appropriate regularizations, enabling convergence to the desired locality-sensitive property. By doing so, {\sc DeepLSH} can mimic any given similarity metric, thereby enhancing and improving the time and efficiency of near-duplicate crash report detection. Overall, our findings highlight the untapped potential of LSH in the crash bucketing domain. We present {\sc DeepLSH} as a practical solution that effectively improves the time and efficiency of automatic crash bucketing. Furthermore, {\sc DeepLSH} maintains compatibility with various similarity metrics, making it a versatile tool for software developers.

\balance
\bibliographystyle{ACM-Reference-Format}
\bibliography{references}
\newpage
\appendix
\end{document}